\documentclass{aa}
\usepackage{txfonts}
\usepackage{psfig}
\begin{document}

\newcommand{\beqa}{\begin{eqnarray*}}
\newcommand{\eeqa}{\end{eqnarray*}}
\newcommand{\beqan}{\begin{eqnarray}}
\newcommand{\eeqan}[1]{\label{#1}\end{eqnarray}}
\newcommand{\beq}{\begin{equation}}
\newcommand{\eeq}{\end{equation}}
\newcommand{\diff}{{\rm d}}
\newcommand{\drr}{\frac{\partial}{\partial r}}
\newcommand{\dtt}{\frac{\diff}{\diff t}}
\newcommand{\dr}[1]{\frac{\partial  #1}{\partial r}}
\newcommand{\dt}[1]{\frac{\partial  #1}{\partial t}}
\newcommand{\lp}{ \left(}
\newcommand{\rp}{ \right)}
\newcommand{\lc}{ \left[}
\newcommand{\rc}{ \right]}
\newcommand{\cf}{{\it cf.}~}
\newcommand{\ie}{{\it i.e.}~}
\newcommand{\eg}{{\it e.g.}~}

\def\O{\Omega}
\def\lta{\mathrel{{\lower 3pt\hbox{$\mathchar"218$}} \hspace{-8pt}
     \raise 2.0pt\hbox{$\mathchar"13C$} \hspace{-5pt} }}
\def\gta{\mathrel{{\lower 3pt\hbox{$\mathchar"218$}} \hspace{-8pt}
     \raise 2.0pt\hbox{$\mathchar"13E$} \hspace{-5pt} }}

\bibliographystyle{plain}

\title{Angular Momentum Transport by Internal Gravity Waves}
\subtitle{I - Pop I Main Sequence Stars}

\author{Suzanne Talon\inst{1,2} and Corinne Charbonnel\inst{3,4}}

\offprints{Suzanne Talon}

\institute{
CERCA, 5160, boul. D\'ecarie, suite 400, Montr\'eal PQ H3X 2H9
\and
D\'epartement de Physique, Universit\'e de Montr\'eal, Montr\'eal PQ H3C 3J7
\and Laboratoire d'Astrophysique de Toulouse, CNRS UMR 5572, OMP,
14, Av. E.Belin, 31400 Toulouse, France 
\and Observatoire de Gen\`eve, 51, ch. des Maillettes, 1290 Sauverny, Switzerland \\
(Suzanne.Talon@cerca.umontreal.ca, Corinne.Charbonnel@obs.unige.ch)}

\date{Received 28 January 2003 / Accepted 16 April 2003}

\authorrunning{S. Talon \& C. Charbonnel}
\titlerunning{Gravity waves in Pop I main sequence stars}

\abstract{
We examine the generation of gravity waves by the surface convection
zone of low-mass main sequence stars with solar metallicity. 
It is found that the total momentum luminosity in waves
rises with stellar mass, up to the quasi-disappearance of the convection
zone around 6500\,K (corresponding to a mass of $\sim$1.4 $M_\odot$ for
solar metallicity) where the luminosity drastically drops.
We calculate the net momentum extraction associated with these waves and
explain how the calculated mass dependence helps resolve the enigma of the
Li dip in terms of rotational mixing, forming a coherent picture of
mixing in all main sequence stars.
\keywords{Hydrodynamics;  
Stars: interiors, late-type, rotation; Turbulence; Waves}
}

\maketitle

\section{Clues to angular momentum transport in low-mass stars}

In many locations of the Hertzsprung-Russell diagram, stars exhibit
signatures of processes that require challenging modeling beyond the 
standard stellar theory\footnote{By standard we refer to the modeling 
of non-rotating, non-magnetic stars, in which convection is the only large-scale
mixing considered.}.
In this context, rotation has become a major ingredient of modern models,
especially when abundance anomalies have to be accounted for. 
In order to correctly describe the effects of rotation on stellar structure and
evolution as well as on their ``byproducts'' like surface abundances or chemical yields,
special emphasis has to be put on the 
evolution of the angular momentum distribution;
this indeed is the feature that determines the extent 
and magnitude of rotation-induced mixing in stars.

In recent theoretical developments, 
the internal rotation law evolves as a result of contraction, expansion,
meridional circulation, shear turbulence and mass loss; 
mixing of chemicals is directly linked to the rotation profile 
(see e.g Talon 2003). 
The most sophisticated
treatment of these hydrodynamical processes is based mainly on 
the work by Zahn (1992), Maeder (1995), Talon \& Zahn (1997) and Maeder \& Zahn (1998).
It rests on only one assumption which is that the turbulence 
sustained by the shear is highly anisotropic. 
Two free parameters describe the magnitude of the horizontal shear 
and the erosion of the restoring force due to both the thermal and 
the mean molecular weight stratifications. 

Such a self-consistent treatment has been successfully applied in various parts 
of the HR diagram. 
The strength of the theory is that the use of the same free parameters give 
very satisfactory results for stars over a large range of masses and 
evolutionary phases. For example, it explains the observed He and N enrichment 
in main sequence O-type and early B-type stars, in OB supergiants 
(e.g., Gies \& Lambert 1992, Lennon et al. 1991, Herrero et al. 1999) 
and in A-type supergiants in the SMC (Venn 1999), 
as well as the B depletion in main sequence B-type stars (e.g., Venn et al. 1996); 
it also helps reproducing the number ratio of blue to red supergiants in the 
SMC (e.g., Meylan \& Maeder 1982) and the observed WR/O ratios at solar metallicity.
We refer to Maeder \& Meynet (2000) for a more detailed description of the 
effects of rotation in massive stars and for relevant references.

In low-mass stars, the strongest observational constraints come 
from the so-called Li dip and from helioseismology. 
The Li dip is a characteristic feature of lithium abundances 
which is seen both in the field and in open clusters 
(e.g., Wallerstein et al. 1965, Boesgaard \& Tripicco 1986, Balachandran 1995). 
It corresponds to a narrow region in effective temperature (around $T_{\rm eff} \sim~$6650\,K) 
where the surface Li abundances are reduced by up to $\sim$2.5 dex. 
In the same region, 
low-mass stars are beginning to be spun down efficiently
early on the main sequence (Boesgaard 1987), most probably via magnetic torquing. 
In modern rotating stellar models in which the transport of angular momentum and 
of chemicals by meridional circulation and shear turbulence is self-consistently 
taken into account, this leads to important lithium (and beryllium) destruction 
at the correct $T_{\rm eff}$ (Talon \& Charbonnel 1998, hereafter TC98; 
Palacios et al. 2003, hereafter PTCF03). 
In addition rotation induced mixing inhibits the atomic diffusion 
in a way that explains the constancy of the CNO abundances within the Li dip 
(e.g., Varenne \& Monier 1998, Takeda et al. 1998). 
Last but not least, these models do explain the evolution of Li surface abundances
in evolved stars that originate from the hot side of the dip. 
It is worth recalling that all these results for low-mass stars on the blue side of the dip 
are obtained with the same values for the two 
above-mentioned parameters as those used in massive stellar models.

For still cooler main sequence stars however,
the magnetic torque strengthens as the stellar convective envelope grows. 
If we assume here that all the momentum transport is assured by the 
wind-driven meridional circulation as it is on the blue side of the Li-dip, 
then too much lithium burning is obtained below $T_{\rm eff} \sim 6550$\,K.
Thus the rise of the Li abundance on the red side of the dip can be interpreted 
as the signature of another mechanism that efficiently transports 
angular momentum (but not chemicals) in lower mass stars (TC98). 
As a result, in these objects the magnitude of both meridional circulation and 
shear turbulence is reduced, as well as the Li depletion due to rotational mixing.

Such a mechanism is also required to shape the Sun's flat rotation profile 
(Chaboyer et al. 1995, Matias \& Zahn 1998). At the solar age, 
models relying solely on turbulence and meridional circulation for
momentum transport
still predict large angular velocity gradients which are not present in the Sun.

Only two mechanisms have been proposed so far to transport angular momentum 
in addition to the classical hydrodynamical processes in order to enforce the 
flat rotation profile measured by helioseismology :
magnetic fields (Charbonneau \& MacGregor 1993, Barnes et al. 1999) and gravity waves
(Schatzman 1993, Kumar \& Quataert 1997, Zahn, Talon \& Matias 1997). 
As explained in TC98, in order to obtain a consistent picture 
of rotational mixing in all
types of stars, the correct mechanism must become efficient
in the center of the Li dip, namely at an effective temperature
of $\sim~$6650\,K.

While the original propositions for momentum extraction by gravity waves 
have received much criticism (Gough \& McIntyre 1998, Ringot 1998),
recent calculations performed by Talon, Kumar \& Zahn (2002, hereafter TKZ02) 
show that, through differential filtering, gravity waves are indeed able to extract
momentum from the solar interior.
Here we explore the same physics in the stellar mass range around the Li dip
in order to verify  whether it has the proper mass dependence 
as described by TC98 (see Fig.~1).
Our goal is to present a coherent understanding of rotational
mixing in {\em all} stars of the main sequence. 
We will firstly recall the main descriptions that have been
made for wave excitation (\S~2) and explain how thermal
dissipation of these waves leads to the formation of
a filtering shear layer (\S~3). Results for momentum extraction
by waves in stars of various masses are then presented (\S~4)
and discussed (\S~5).

\begin{figure}
\centerline{
\psfig{figure=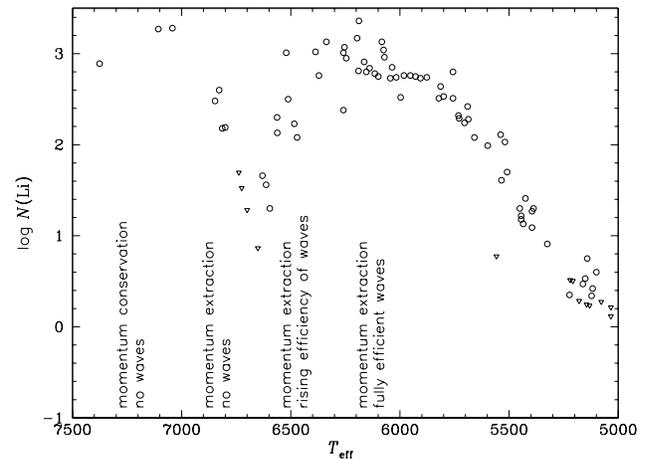,height=6.5cm,angle=-90}
}
\caption{Measured Li abundances in the Hyades. Superposed is the 
pattern of efficiency of momentum transport required
to produce the Li gap via rotational mixing
}
\end{figure}

\begin{table*}
\caption{Characteristics of the stellar models (with $Z=0.02$) on the ZAMS 
and maximum frequency $\nu_{\rm max}$ used to capture the wave - mean flow interaction.
$\ell_c$ is the largest convective scale,  $\nu_c$, the characteristic convective frequency
and $K_T$, the thermal diffusivity.}
\begin{center}
\begin{tabular}{ccccccccc}
\hline
\multicolumn{1}{c}{$M_*$}
& \multicolumn{1}{c}{$T_{\rm eff}$}
& \multicolumn{1}{c}{$T_{\rm eff}$} 
& \multicolumn{1}{c}{log$\left({M_{\rm zc}/{M_*}}\right)$}
& \multicolumn{1}{c}{$\ell_c$} 
& \multicolumn{1}{c}{$\nu_c$}
& \multicolumn{1}{c}{$K_T$}
& \multicolumn{1}{c}{$\nu_{\rm max}$}
\\
(M$_{\odot}$) & (K) & (K)&  &  & ($\mu$Hz) & 
(cm$^2$~s$^{-1}$) & ($\mu$Hz)\\
 & zams & Hyades \\
\hline
1.1  & 5615 & 5640 & $-2.26$ & 80   & 1.28 &  $4.49 \times 10^{7}$ & 5.25 \\
1.2  & 5905 & 5950 & $-3.09$ & 117  & 3.30 &  $4.49 \times 10^{8}$ & 5.25 \\
1.3  & 6210 & 6260 & $-4.12$ & 184  & 9.34 &  $5.58 \times 10^{9}$ & 5.25 \\
1.35 & 6370 & 6420 & $-4.92$ & 269  & 22.8 &  $3.06 \times 10^{10}$ & 10.5 \\
1.4  & 6555 & 6595 & $-5.77$ & 407  & 54.2 &  $1.75 \times 10^{11}$ & 15.5 \\
1.42 & 6620 & 6670 & $-6.44$ & 599  & 114  &  $5.47 \times 10^{11}$ & 15.5 \\
1.43 & 6665 & 6710 & $-6.84$ & 754  & 182  &  $1.12 \times 10^{12}$ & 15.5 \\
1.44 & 6705 & 6750 & $-7.13$ & 892  & 264  &  $1.96 \times 10^{12}$ & 16.5 \\
1.45 & 6750 & 6785 & $-7.34$ & 1045 & 371  & $ 2.26 \times 10^{12}$ & 18.0 \\
\hline
\end{tabular}
\end{center}
\end{table*}

\section{Turbulent wave excitation}

\subsection{Two excitation mechanisms}

The generation of gravity waves depends on the structure
of the stellar convective envelope. 
On the main sequence, this region is rather large in solar mass stars
($\sim$30$\%$ in radius), while it 
shrinks as the stellar
mass rises, representing only $\sim$1$\%$ in
a $1.45~M_\odot$ star lying inside
the Li dip. The characteristic convective frequency
varies from $\nu_c \sim$0.75~$\mu$Hz to $\nu_c \sim$600~$\mu$Hz while the spherical
harmonic number $\ell$ corresponding to the largest convective scale varies
from $\ell_c \sim$60 to $\ell_c \sim$1000 in the same mass range.
These variations are not smooth, rapid 
changes of convective
properties occurring when 
$M_*\gta\,1.3~M_\odot$. 
We wish to investigate how the wave spectrum 
produced by these convection zones
varies as a function of stellar mass.

The exact properties of wave spectra
remain somewhat uncertain. Excitation can be related
to internal stresses that correlate with the mode's eigenfunction
(Goldreich \& Kumar 1990, Balmforth 1992, Goldreich et al. 1994).
Gravity waves may also be excited by
penetration below a convection zone, as observed in laboratory
experiments (Townsend 1958) and in 
numerical simulations (Hurlburt et al. 1986, 1994, Nordlund et al. 1996,
Kiraga et al. 2000).
In a real star, both sources would contribute to wave generation
and they are thus additive.

In this paper, we examine both sources, with emphasis not on
their absolute values, but rather on their variation with 
stellar mass.
Our computations are performed for stars between $1.1$ and 
$1.45~M_{\odot}$ with $Z=0.02$
(see Table 1). We use the same code and input physics 
(eos, opacities, nuclear reactions) as in PTCF03. Wave characteristics 
are computed on the basis of ZAMS stellar structures.
Note that in these objects the depth and structure of the convective 
envelope does not vary significantly over a main sequence lifetime
(at least up to the age of the Hyades).

\subsection{Excitation by the Reynolds stresses}

First, we consider excitation by the Reynolds stresses as described
by Goldreich et al. (1994). This description, which has also been
used by Kumar, Talon \& Zahn (1999, hereafter KTZ99), 
uses a free
parameter that has been calibrated on the solar {\bf p}-mode
spectrum. 

The energy flux per unit frequency due to Reynolds stresses
${{\cal F}_E}^R$ is then
\begin{eqnarray}
{{\cal F}_E}^R \lp \ell, \omega \rp &=& \frac{\omega^2}{4\pi} \int _{r_c}^R dr\; \frac{\rho^2}{r^2} 
   \left[\left(\frac{\partial \xi_r}{\partial r}\right)^2 + 
   \ell(\ell+1)\left(\frac{\partial \xi_h}{\partial r}\right)^2 \right]  \nonumber \\
 && \times  \exp\left[ -h_\omega^2 \ell(\ell+1)/2r^2\right] \frac{v^3 L^4 }{1 
  + (\omega \tau_L)^{15/2}},
\label{gold}
\end{eqnarray}
where 
$\xi_r$ and $[\ell(\ell+1)]^{1/2}\xi_h$ are the radial and horizontal
displacement wave-functions which are normalized to unit energy flux just 
below the convection zone, $v$ is the mixing length 
convective velocity, $L$ is the radial
size of an energy bearing turbulent eddy, $\tau_L \approx L/v$ is the
characteristic convective time, and $h_\omega$ is the
radial size of the largest eddy at $r$ with characteristic frequency of
$\omega$ or greater ($h_\omega = L \min\{1, (2\omega\tau_L)^{-3/2}\}$).
The gravity waves are evanescent in the convection zone, the region
where they are excited. Their wave-functions $\xi_r$ and $\xi_h$ are thus proportional
to $k_r^{-1/2} \, \exp \lp i\int \diff r \, k_r \rp$.
The above equation was derived under the assumption that the
turbulence spectrum is Kolmogorov.
The momentum flux per unit frequency ${{\cal F}_J}^R$ is then
\begin{equation}
{{\cal F}_J}^R \lp m, \ell, \omega \rp = \frac{m}{\omega}  {{\cal F}_E}^R \lp \ell, \omega \rp.
\label{mom_flux}
\end{equation}
The momentum spectra corresponding to the various stellar
masses are shown in Fig.~2 where 
${\cal L}_J=4\pi r_{\rm cz}^2 {\cal F}_J$ 
is the momentum luminosity; note that some of the
lower frequency waves have zero magnitude. This is due to the fact that the
corresponding damping is too large to permit them to be waves.

Spectrum characteristics evolve together with the structure of the
convection zone which depends on the stellar mass.
Low frequency waves disappear when the stellar
mass increases. This is related to the fact that 
the surface convection zone
then becomes thinner, leading to a larger thermal diffusivity just
below it. This in turn implies stronger damping (see Eq.~\ref{optdepth}) and
the disappearance of low frequency waves. 
Furthermore, as stellar mass rises
up to about $1.35~M_{\odot}$ (i.e., $T_{\rm eff} \sim $6400~K), 
so does the flux associated with
a given frequency. This is related to the rise of the luminosity with mass,
and thus, of the energy in convective motions. 
For even more massive stars, the flux associated with a given frequency
remains about constant, but most low frequency waves are damped too
rapidly to exist.

Let us recall that these results are obtained for stellar 
structures on the ZAMS,
and that we have found no significant dependence of the wave spectrum with
age on the main sequence; it can thus be considered
as constant for the duration of the main sequence.

In this description, waves with frequencies up to $N_c$ (the Brunt-V\"ais\"al\"a
frequency at the base of the convective envelope) 
can be excited. However, the high-frequency waves will have little impact on 
momentum evolution; indeed, as frequencies increase, filtering by
the shear layer 
(see \S 3) becomes less efficient and less differential. Furthermore,
damping is so small that 
it leads to only a small amount of momentum
redistribution. Finally, as wave excitation diminishes with frequency, they
carry much less momentum overall. 
These high-frequency waves will form standing waves, the {\bf g}-modes
of helio- and astero-seismology.

\begin{figure*}[t]
\centerline{
\psfig{figure=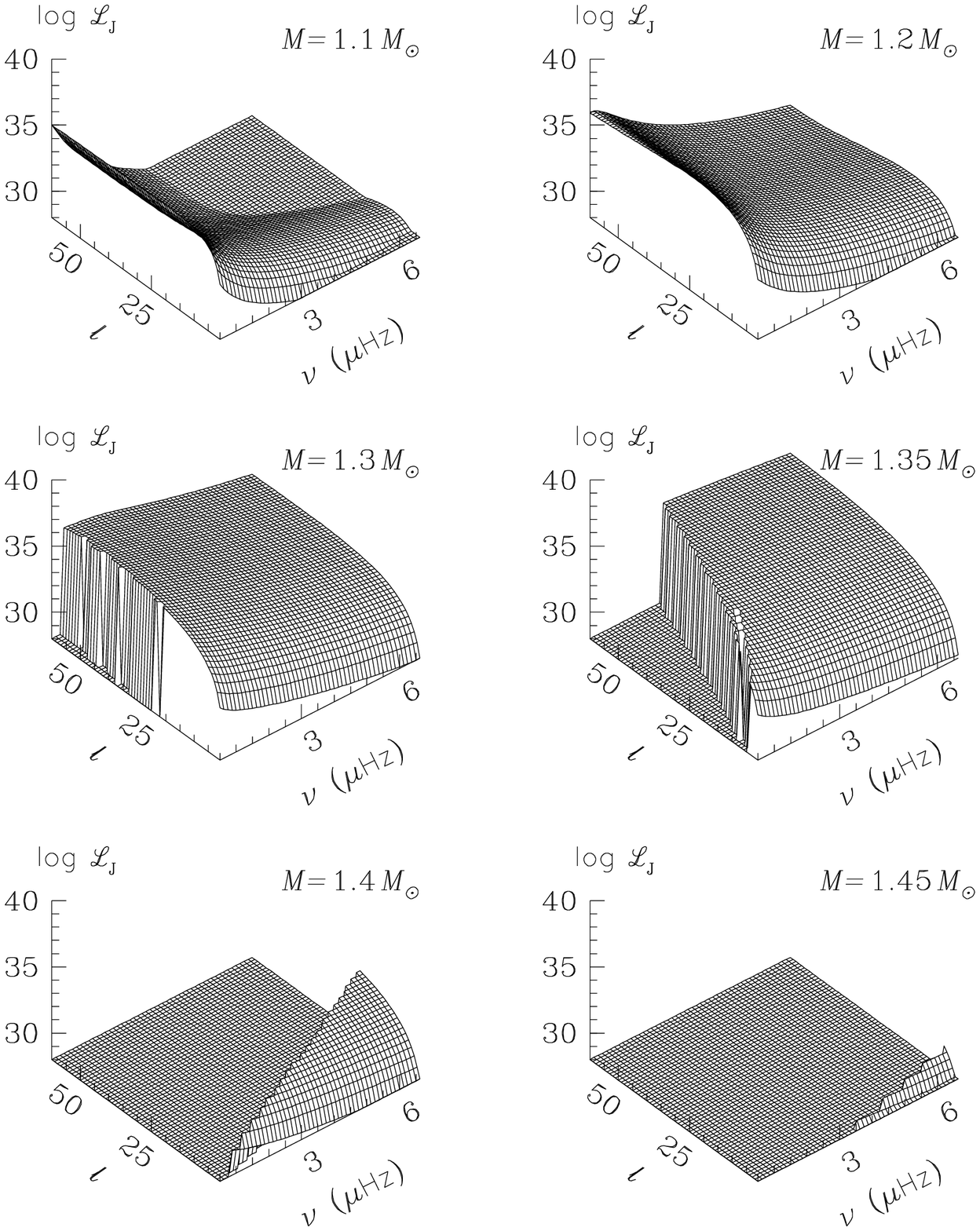,height=12cm}
\psfig{figure=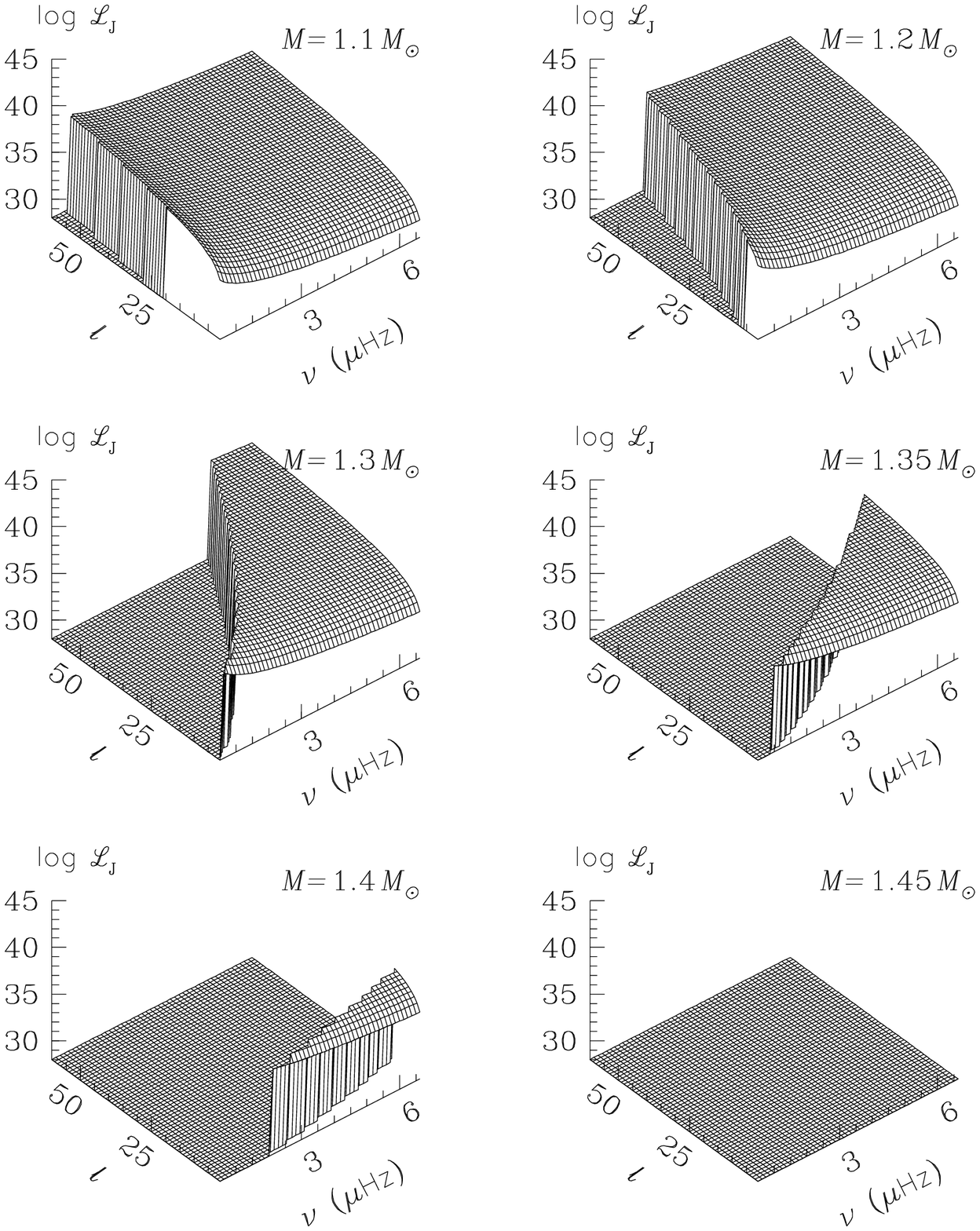,height=12cm}
}
\caption{Momentum spectrum in $\ell$ and $\nu$ for various masses 
for {\bf (left)} Goldreich, Murray, \& Kumar model
and {\bf (right)} Garc\'\i a L\'opez \& Spruit model
\label{specgold}}
\end{figure*}

\subsection{Excitation by convective penetration}

Greater uncertainties exist on wave generation via convective
penetration. The only theoretical 
estimates that exist have been
made by Press (1981) and Garc\'\i a L\'opez \& Spruit (1991), and later
used by Zahn et al. (1997). It is the formulation we shall adopt 
here.

This description is based on the matching of wave pressure fluctuations
with those of turbulent convection. Furthermore, it takes into
account the combination of turbulent eddies of a given size to produce
larger fluctuations. The range of horizontal scales available is thus
\beq
0 < \ell < \ell _u ~~~~~ 
{\rm with} ~ \ell_u = \ell _c \lp \frac{\omega}{\omega_c} \rp ^{3/2}
\eeq
where $\ell_c$ is the spherical harmonic number associated with the
largest convective scale and $\omega_c$ is the corresponding convective frequency.
Turbulence is assumed to follow a Kolmogorov spectrum, all
frequencies with $\omega \ge \omega _c$ thus being available.
The associated energy flux ${{\cal F}_E}^P$ is
\beq
{{\cal F}_E}^P \lp \ell, \omega \rp = \frac{\rho v_c^3}{2} \, \frac{\omega_c^3}{\omega^2}\,
\frac{\ell}{\ell_c} \, \frac{1}{N} \, \lp 1- \frac{\omega^2}{N^2} \rp ^2
\eeq
and the corresponding angular momentum flux is still given by Eq.~(\ref{mom_flux}).

\subsection{A word of caution}
Let us stress 
that the fluxes calculated here are somewhat uncertain,
especially in the second case. Indeed, it is well known that the ``structure''
of convection, 
which contains e.g. plumes
that travel down across the whole convection
zone, is not well reproduced by the mixing length model even though the
convective velocities are more or less correct (e.g. Hurlburt et al. 1986). However, their
differential properties should have the proper dependence, 
and this will be our focus in the rest of this paper.

\begin{figure}[t]
\centerline {
\psfig{figure=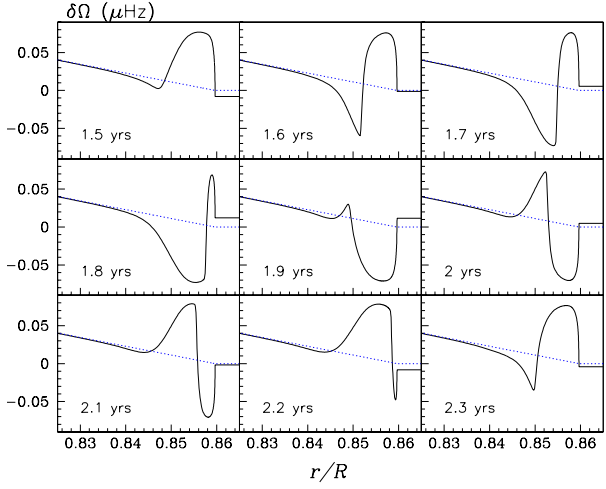,height=6.5cm,angle=-90}
}
\caption{Oscillating shear layer below the surface convection zone
of a $1.2~M_\odot$ star.
The dotted line shows the initial rotation profile.
With the surface rotating slower than the core, a prograde shear
layer is initially formed, followed by a retrograde one. When the
shear becomes too intense, turbulent viscosity acts to merge the
prograde layer with the convection zone, leaving behind the retrograde layer. 
A new prograde layer forms behind, and migrates towards the convection zone when the
retrograde layer is  absorbed. The cycle then resumes.
\label{fig:local}}
\end{figure}

\section{Formation of a shear layer}

\subsection{Differential filtering of the waves by a shear layer}

As described by several authors (Gough \& McIntyre 1998, Ringot 1998,
KTZ99, Kim \& MacGregor 2001), the most obvious feature of momentum transport by gravity
waves is the formation of a double peaked shear layer just below
the convection zone (see Fig.~\ref{fig:local}).

Gough \& McIntyre (1998) and Ringot (1998) argued that such a layer would
prevent waves from propagating beyond. Indeed, local radiative and viscous
damping is greatly dependent on the local frequency and 
increases
when that frequency diminishes (this property leads to the formation
of the double peaked shear layer).
The local momentum luminosity integrated over the whole spectrum writes
\begin{equation}
{\cal L}_J(r) = \sum_{\sigma, \ell, m} {{\cal L}_J}_{\ell, m} \lp r_{\rm zc}\rp
\exp \lc -\tau(r, \sigma, \ell) \rc.
\end{equation}
The local amplitude 
$\exp \left[ -\tau(r, \sigma, \ell)\right]$ depends on the 
integrated damping due to thermal diffusion $K_T$ and (turbulent) viscosity $\nu_t$
\begin{eqnarray}
\lefteqn{\tau(r, \sigma, \ell) = }  \label{optdepth} \\
&&[\ell(\ell+1)]^{3\over2} \int_r^{r_c} 
\lp K_T + \nu_t \rp \; {N N_T^2 \over
\sigma^4}  \left({N^2 \over N^2 - \sigma^2}\right)^{1 \over 2} {\diff r
\over r^3} \nonumber
\end{eqnarray}
where $N^2 = N_T^2 + N_{\mu}^2$ is the Brunt-V\"ais\"al\"a frequency, 
$N_T^2$ is its thermal part 
and $ N_{\mu}^2$ is due to the mean molecular weight stratification
(Zahn et al. 1997).

The prograde peak\footnote{The prograde peak 
corresponds to the layer 
that rotates
more rapidly than the convection zone, while the retrograde 
peak designates the layer that rotates more
slowly.} thus filters prograde waves, while the retrograde peak filters retrograde
waves. If the peaks were infinite in height, not a single wave could
travel through it. This is however not the case, as the magnitude of the shear layer
is self regulated by shear 
turbulence\footnote{Baroclinic instabilities could
also set in, and would have the same effect.}.

As described by TKZ02, if no initial differential rotation is present,
the average wave momentum luminosity that traverses the shear layer is
null, and there is no net effect on the interior. However, if differential
rotation is initially present, as in the case of low mass stars that are braked by
a magnetic torque, the average magnitude of the two peaks is not equal;
in the case of interest ({\it i.e.} with the surface convection zone
rotating more slowly than the 
interior), the prograde peak is always larger, leading to
differential filtering that favors the penetration of retrograde waves.
These waves are then damped in the whole radiative region
that they spin down by depositing negative momentum.

\subsection{Dynamics of the shear layer}

To study the effect of gravity waves in
low mass stars, one must thus be able not only to evaluate the 
global momentum luminosity in waves, but also the dynamics of the shear layer and the
resulting differential filtering. This is addressed here by performing
numerical simulations of angular momentum evolution. 
We study a few cycles and evaluate the resulting net momentum luminosity for
various masses and various differential rotations.

Angular momentum evolves under the action of gravity waves and shear turbulence
according to
\begin{equation}
\rho \dtt \lc r^2 {\Omega}\rc = \frac{1}{ r^2} \drr \lc \rho \nu_t r^4 \dr{\Omega} \rc 
- \frac{3}{8\pi} \frac{1}{r^2} \drr{{\cal L}_J(r)} 
\label{ev_omega}
\end{equation}
where $\rho$ is the density, $\Omega$ the angular velocity and
$\nu_t$ the turbulent viscosity.
The local momentum luminosity ${\cal L}_J(r)$ is integrated over
the whole wave spectrum.
Equation (7) is solved throughout the radiation zone.
The upper boundary condition expresses conservation of
momentum of the star as a whole, with the convection zone rotating 
as a solid body. 
On their way to the core, waves are reflected when their
local frequency $\sigma$ equals the Brunt-V\"ais\"al\"a frequency 
\begin{equation}
\sigma(r,m) \equiv \omega - m \lc \Omega(r)-\Omega_c \rc = N.
\end{equation}
Waves then deposit more momentum as they travel back to the surface convection zone.
The high-frequency waves of lowest degree considered in this study
can be reflected up to 5000 times
before they are completely damped; these contribute little to momentum
redistribution\footnote{As the frequency rises, damping diminishes and standing
waves may form.}.

\begin{figure}[t]
\centerline{
\psfig{figure=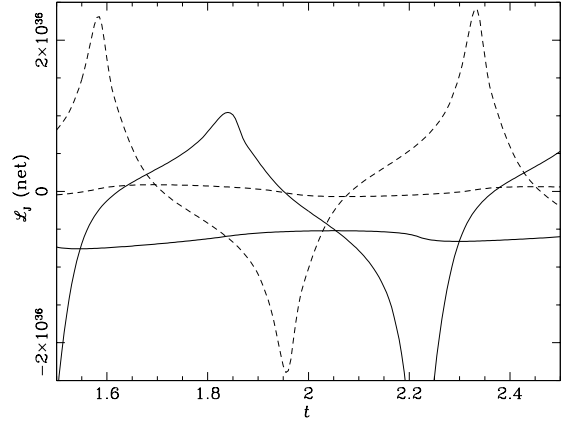,height=6cm,angle=-90}
}
\caption{Net luminosity vs time (in years)
at $0.03\,R_*$ below the surface convection zone.  
Solid lines correspond to the case of strong initial differential
rotation ($\delta \Omega = 0.05\,\mu {\rm Hz}$ over $0.05\,R_*$) and
dashed lines to the case of small initial differential
rotation ($\delta \Omega = 0.00005 \,\mu {\rm Hz}$ over $0.05\,R_*$).
Bold lines represent the mean luminosity, thin lines represent the
instantaneous luminosity.
The curves correspond to a $1.2~M_\odot$ model.
The shear layer oscillation occurs here with a period of $\sim 1$ year.
\label{netflux}}
\end{figure}

\begin{figure}[t]
\centerline{
\psfig{figure=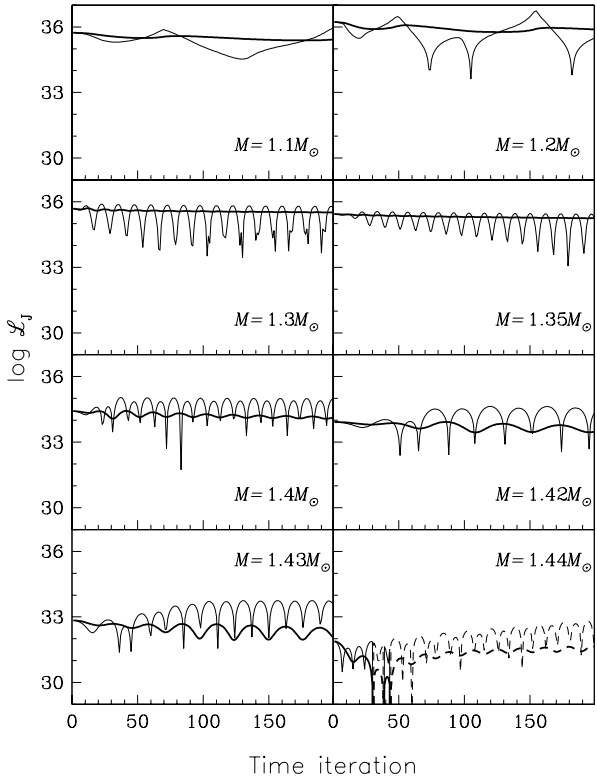,height=11cm}
}
\caption{Net average (bold) and instantaneous (thin) luminosities $0.03\,R_*$ below the surface convection zone
after a few cycles for various masses and in the case of ``strong'' initial
differential rotation ($\delta \Omega = 0.05\,\mu {\rm Hz}$ over $0.05\,R_*$).
The dashed line corresponds to positives luminosities.
\label{compflux}}
\end{figure}

\begin{figure}[t]
\centerline{
\psfig{figure=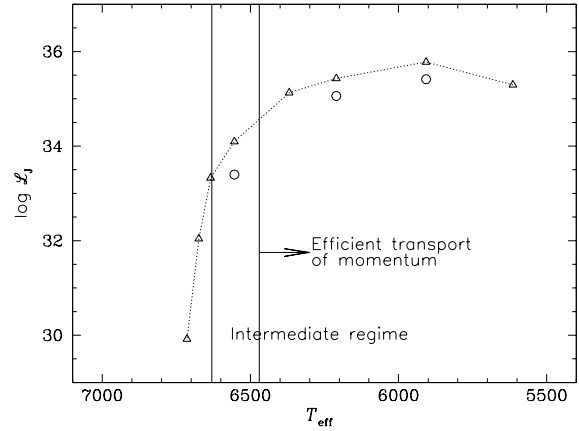,height=6cm,angle=-90}
}
\caption{Net luminosity at $0.03\,R_*$ below the surface convection zone
as a function of $T_{\rm eff}$(zams) for an initial differential
rotation of $\delta \Omega = 0.05\,\mu {\rm Hz}$ over $0.05\,R_*$.
Triangles: without the Coriolis force; circles: including the Coriolis force.
The vertical lines correspond to the requirement of angular momentum
efficiency as described by TC98 (they correspond to the horizontal
lines of their Fig.~2).
\label{evoLJ}}
\end{figure}

Figure~3 shows the oscillation cycle of the shear layer for our $1.2~M_{\odot}$ star. 
Each 
model (i.e., each stellar mass) has a different characteristic time-scale for this oscillation
which depends mainly on the net wave flux in the high degree, low frequency
wave.
It is thus shortest for masses around $1.2 ~ - ~ 1.3~M_\odot$.\footnote{Let us note
that the very different time-scale found in TKZ02 for the Sun is related not only
to the smaller flux but also to the reduced number of frequencies and degrees
used in order to be able to perform the 
solar calculation over the required $10^7$
time-steps.}

\section{Momentum extraction}
\subsection{Fluxes}

In all our stellar models, we follow the detailed time evolution of the shear layer
according to Eq.~(\ref{ev_omega}) over several cycles, for two values of the initial
differential rotation (a very small differential rotation of 
$\delta \Omega = 0.00005\,\mu {\rm Hz}$ over $0.05\,R_*$, and a larger differential
rotation $\delta \Omega = 0.05\,\mu {\rm Hz}$ over $0.05\,R_*$). 
We calculate the net luminosity below the filtering shear
layer, at $r=R_{\rm cz} - 0.03 R_*$.
Resulting instantaneous and mean luminosities are detailed for the $1.2~M_\odot$ model 
in Fig.~\ref{netflux}. 
The change of sign of
the instantaneous luminosity illustrates the varying asymmetry of the double peak shear
layer. In the case of very small initial differential rotation, the curve
is symmetric and the mean luminosity is always close to zero. On the other hand, if
the initial differential rotation is large, retrograde waves are favored, as illustrated
by the negative mean luminosity.
It is this net mean luminosity below the shear layer that gives rise to a net momentum
extraction in the interior.

In order to obtain differential filtering, it is important to include
several different modes;
\begin{itemize}
\item High-$\ell$ modes, that damp close to the convection zone
(see Eq.~\ref{optdepth});
\item Low-$\ell$ modes, that penetrate deeply.
\end{itemize}
Furthermore, it is the lowest frequency waves (which undergo
the largest amount of differential damping) that give rise
to momentum extraction. While wave frequencies range from
$\sim 0$ to $N$, for each considered mass, there is a maximum
frequency for waves that contribute to differential filtering.
That maximum frequency has been determined for each individual
mass
(see Table~1). In the most massive models considered here, which correspond
to stars 
inside the dip, 
this maximum frequency is much higher than
in lower mass stars. Indeed, as can be seen from the wave spectrum
(Fig.~\ref{specgold}), in those more massive models,
high-$\ell$ modes of low frequency experience too much damping
to travel beyond a single vertical 
wavelength in the vertical.
One must thus include higher wave frequencies in order to capture
high-$\ell$ modes that produce the oscillating double peaked shear
layer. 

The Garc\'{\i}a L\'opez \& Spruit model for wave excitation by
convective penetration leads to the generation of larger
frequency waves, leading to a limited amount of extraction. This is
especially true for the 
more massive stars presented in this study and is thus less relevant to
our discussion here\footnote{Let us remind that both generating sources
are actually additive.}. For that reason, that model has been ignored in
detailed calculations. 

The time evolution of the resulting filtering in all stellar masses considered 
is shown in Fig.~\ref{compflux}. The abscissa represents the time iteration,
each mass having its own characteristic time-scale for oscillation. 
The results are thus shown in terms of this characteristic time, and not in a
real physical time sequence. 
The net luminosity slightly increases with the stellar mass up to $1.2~M_{\odot}$, and 
then dramatically decreases as one moves to stellar masses inside the Li dip.
For the largest mass considered here 
($1.44$ and $1.45~M_\odot$),
the process of differential filtering does not lead
to momentum extraction anymore 
(the net luminosity is actually positive although very small for 
these stars on the blue side of the dip\footnote{This reverse effect 
is probably
linked to the fact that thermal damping dominates just below the
convection zone. Then, the low-$\ell$, high frequency waves that
are included and that travel back and forth over 100 times, 
are damped mostly there. However, retrograde waves
that have a larger frequency in the interior are less damped there
than the prograde waves, leading to an effect opposite to what
is observed in other stars.}).

Figure~\ref{evoLJ} represents the average luminosity after 500 time steps
as a function of effective temperature. On the same figure are shown
the limits established by TC98 in terms of momentum extraction 
required in order to consistently understand the Li dip in terms
of rotational mixing. 

\subsection{Coriolis force}

\begin{table}
\caption{Effect of the Coriolis force. \label{coriolis}}
\begin{center}
\begin{tabular}{ccccccccc}
\hline
\multicolumn{1}{c}{$M_*$}
& \multicolumn{1}{c}{$T_{\rm eff}$}
& \multicolumn{1}{c}{$\left< v \right> _{\rm Hyades}$}
& \multicolumn{1}{c}{$\nu_{\rm rot}$}
& \multicolumn{1}{c}{$\nu_{\rm wave}$}
& \multicolumn{1}{c}{$\theta_{\rm crit}$} 
& \multicolumn{1}{c}{$\cal{S}$}
\\
($M_{\odot}$) & (K) & (km~s$^{-1}$) & ($\mu$Hz) & ($\mu$Hz) &  & (\%)\\
\hline
1.1  & 5640 &  8 & 1.65 &  1  & 72.4 &  30.2 \\
1.2  & 5950 & 10 & 1.90 &  1  & 74.7 &  26.3 \\
1.3  & 6260 & 20 & 3.50 &  1  & 81.8 &  14.3 \\
1.35 & 6420 & 30 & 5.05 &  1  & 84.3 &  9.9 \\
1.4  & 6595 & 50 & 8.10 &  1  & 86.5 &  6.2 \\
1.45 & 6790 & 75 & 11.5 &  1  & 87.5 &  4.3 \\
\hline
\end{tabular}
\end{center}
\end{table}

In order to fully describe wave transport in rotating stars, we must
also take into account the influence of the Coriolis force on waves.
Its first order effect is to change the horizontal structure of the modes, 
confining them closer to the equator. The maximum co-latitude of
propagation $\theta_{\rm crit}$ is given by the condition
\beq
\sigma ^2 = 4 \Omega ^2 \cos ^2 \theta_{\rm crit}
\eeq
(KTZ99). For a given rotating velocity, this condition implies
that only a fraction
\beq
{\cal{S}} = \int _{\theta_{\rm crit}} ^{\pi / 2} \sin \theta ~\diff \theta
\eeq
of the surface will support waves and the corresponding 
luminosity should be diminished accordingly. Table~\ref{coriolis} gives
the critical angle for propagation and the corresponding efficiency
for a typical frequency of $1~\mu$Hz. The rotation velocities used here 
correspond, for each stellar mass, to the mean velocity observed 
at the age of the Hyades.
Let us note that the fraction of the surface $\cal{S}$ varies linearly
with $\nu$, and that the whole surface will support waves when
$\nu \ge 2~\nu_{\rm rot}$.

Exact calculations can be made by reducing the wave luminosity
corresponding to each frequency by the proper coefficient. This has been
done for 3 models ($1.2$, $1.3$ and $1.4~M_\odot$).
The amplitude of low frequency waves (that dominate the formation of the shear layer)
being reduced, including the Coriolis force leads to a somewhat smaller
double peaked shear layer, that oscillates on a slightly longer period.
Furthermore, differential filtering corresponding to momentum extraction
is also dominated by rather low frequency waves. This leads to a smaller
efficiency of momentum extraction by waves (see Fig.~\ref{evoLJ}).
However, the overall mass dependence is not affected.

\section{Discussion}

Several abundance anomalies are best explained in terms of
large-scale mixing. As a natural 
physical parameter (beyond mass,
chemical composition and age), rotation
is an important 
factor in understanding various features which 
are not predicted from classical stellar
models (see Maeder \& Meynet 2000 for details).
Furthermore, if one is to built a consistent model
of rotational mixing, it has to apply to all stars.

Recent improvements of the description of the Eddington-Sweet
meridional circulation by Zahn (1992) and Maeder \& Zahn (1998)
have been applied to several kinds of stars. 
These studies show that, with the same
free parameters, it is possible to explain abundance anomalies
in B (Talon et al. 1997) and O stars (Maeder \& Meynet 2000)
as well as the blue side
of the Li dip (TC98, PTCF03) and Li anomalies
in sub-giants (Charbonnel \& Talon 1999).

However, in order to understand the helioseismic rotation profile, 
some extra mechanism for momentum transport 
is required in low mass stars (Matias \& Zahn 1998).
TC98 argued that, for rotational mixing to be responsible
for the Li dip, this mechanism has to become efficient
at an effective temperature around $\sim 6500-6600\,$K, 
i.e., in a region where the stellar convective envelope 
becomes substantial. 
If the same description is to be applied to all stars, there
must be some physical reason for the lack of efficiency of this
mechanism in hotter stars.
The goal of this paper was to establish whether gravity waves
do have the proper effective temperature dependence.

The calculations we performed here show that the built up of a double
peaked shear layer allowing differential filtering of the waves is
possible in low-mass main sequence stars up to an effective temperature 
of $\sim 6700\,$K. 
The net momentum luminosity below the shear layer increases with stellar mass up 
to $1.2~M_{\odot}$ (i.e., $T_{\rm eff}\sim 5900-5950\,$K); 
it then drops in hotter stars, even becoming slightly positive 
for stars more massive than $\sim 1.44~M_{\odot}$ (i.e., $T_{\rm eff}> 6700\,$K).
While the quantitative results shown
here are based on the excitation model by Goldreich et al. (1994),
the Garc\'{\i}a L\'opez \& Spruit (1991) model would lead to an
even stronger decrease in efficiency. Indeed, as can be seen in 
Fig.~\ref{specgold}, in the more massive models, there are simply
no low frequency waves, implying an even more abrupt decrease
in the efficiency of differential filtering.

Let us discuss a final issue.
As can be seen from Fig.~1, the dispersion in lithium abundances is somewhat larger 
on the cold side of the dip ($T_{\rm eff} \sim 6000 - 6500\,$K) than in cooler 
stars. A very straightworward explanation of this arises within our framework.
As we have shown indeed, for a given stellar mass a higher rotation velocity implies 
a stronger reduction of the net momentum luminosity in waves due to the Coriolis force.
As a consequence, the efficiency of angular momentum extraction by the waves 
is somewhat lower within faster rotators, leading to a larger efficiency of the hydrodynamical 
processes, and thus to stronger Li destruction.  
In this context, the larger lithium dispersion measured in stars laying immediatly 
on the right of the dip can be interpreted as a consequence of the larger dispersion in
rotational velocities that these stars do exhibit in comparison with the less massive ones
(Gaig\'e 1993).
Detailed calculations would have to be performed to check whether this effect could be large enough
to explain the observed dispersion.

{\bf All our results clearly indicate that momentum transport 
by gravity waves has the proper mass dependence to be the required
process in low-mass main sequence stars on the cold side of the Li dip.}

Note that a similar mass dependence is
not expected if momentum transport is dominated
by magnetic fields. There is another test which could permit
to differentiate the magnitude of transport by magnetic field
compared to transport by waves which applies to the Sun.
Certain early helioseismic inversions (Elsworth et al. 1995,
Corbard et al. 1997) suggested that the central region
of the Sun (below $\sim 0.4\,R_\odot$) could rotate somewhat
slower than the rest of the radiative region. These results
seem to be confirmed by recent SOHO data (Couvidat et al. 2003).
TKZ02 explained how such a feature could be produced by
gravity waves.
Both arguments favor momentum transport by gravity waves 
with respect to magnetic field. 

We are now in a position to present a coherent picture of rotational mixing 
in main sequence stars of all masses: \\
On the hot side of the Li dip and in more massive stars, the 
transport of angular momentum and of chemicals by 
meridional circulation and shear instabilities do explain the Li 
as well as the He and CNO patterns. 
In lower mass stars, gravity waves dominate the transport of
angular momentum, 
thereby reducing the magnitude of meridional circulation and shears 
and shaping the Li pattern on the cold side of the dip. 
Within this framework, we thus predict that Pop I main sequence stars 
with initial masses lower than $\sim 1.4~M_{\odot}$ must be quasi-solid 
body rotators, as the Sun is. 

\begin{acknowledgements}
We acknowledge financial support from the 
French Programme National de Physique Stellaire (PNPS).
We thank the R\'eseau qu\'eb\'ecois de calcul de haute performance (RQCHP) 
and the Centre informatique national de l'enseignement sup\'erieur (CINES) 
for useful computational resources.
\end{acknowledgements}

\end{document}